\documentclass[sn-mathphys]{sn-jnl}

\newcommand{\be}{\begin{equation}}
\newcommand{\ee}{\end{equation}}
\newcommand{\bg}{\begin{equation}}
\newcommand{\eg}{\end{equation}}
\newcommand{\bdm}{\begin{displaymath}}
\newcommand{\edm}{\end{displaymath}}
\newcommand{\bea}{\begin{eqnarray}}
\newcommand{\eea}{\end{eqnarray}}
\newcommand{\beas}{\begin{eqnarray*}}
\newcommand{\eeas}{\end{eqnarray*}}
\newcommand{\ba}{\begin{array}}
\newcommand{\ea}{\end{array}}

\newcommand{\bfg}{\begin{figure}}
\newcommand{\efg}{\end{figure}}

\newtheorem{lm}{Lemma}
\newtheorem{cl}{Corollary}
\newtheorem{df}{Definition}

\newcommand{\blm}{\begin{lm}}
\newcommand{\elm}{\end{lm}}
\newcommand{\bcl}{\begin{cl}}
\newcommand{\ecl}{\end{cl}}
\newcommand{\bdf}{\begin{df}}
\newcommand{\brk}{\begin{rm}}
\newcommand{\erk}{\end{rm}}

\newcommand{\ld}{\lambda}

\newcommand{\Gm}{\Gamma}

\newcommand{\ct}{\cite}

\newcommand{\vE}{{\bf E}}

\jyear{2021}%

\begin{document}

\title[Is there a proper figure of merit for a plasmonic structure involved in ...]{Is there a proper figure of merit for a plasmonic structure involved in metal-enhanced fluorescence?}

\author*[1]{\fnm{Ilia L.} \sur{Rasskazov}}\email{irasskaz@ur.rochester.edu}
\author[2]{\fnm{Alexander} \sur{Moroz}}\email{wavescattering@yahoo.com}

\affil*[1]{\orgdiv{The Institute of Optics}, \orgname{University of Rochester}, \orgaddress{
\city{Rochester}, \postcode{14627}, \state{NY}, \country{USA}}}

\affil[2]{Wave-scattering.com}

\abstract{
A suitability of a nanostructure for a metal-enhanced fluorescence (MEF) is usually assessed from the value of the maximal fluorescence enhancement factor, $\mathcal{F}$, which it can generate. 
However, $\mathcal{F}$ is an ambiguous quantity which may dramatically depend on the intrinsic quantum yield, $q_0$, of the emitter in a free space. 
Here we suggest $\mathcal{F}=\mathcal{F}_{q_0=1}$ taken in the limiting case of $q_0=1$ as a proper figure of merit for a plasmonic structure involved in MEF.
Any other realistic $\mathcal{F}$ in the MEF regime can be recovered from $\mathcal{F}_{q_0=1}$ by a simple $1/q_0$ scaling.
}


\maketitle

Plasmonic nanostructures have been considered as a promising tool for enhancing fluorescence emission for a long time~\cite{Ford1984,Geddes2002,Malicka2003,Tovmachenko2006,Anger2006,Bharadwaj2007,Jiao2014,Arruda2017a,Maccaferri2021}.
Collective electron oscillations on the surface of a plasmonic nanostructure can generate a strong local electric field, $\vE$. 
The enhancement of the field boosts the excitation rate, $\gamma_{\rm exc}$ ($\propto \lvert \vE \rvert^2$), of an electric dipole emitter.
At the same time, the presence of an intrinsically dissipative metal nanostructure increases the nonradiative decay rate, $\Gm_{\rm nrad}$,  which diminishes the quantum yield, $q$, and negatively affect fluorescence.
The fluorescence enhancement factor is defined as a product of the excitation rate (at {\em excitation} wavelength, $\ld_{\rm exc}$) and quantum yield (at {\em emission} wavelength, $\ld_{\rm ems}$) of an electric dipole emitter near a plasmonic nanostructure with respect to the same quantities of the emitter in free space:

\begin{equation}
    F = \frac{\gamma_{\rm exc}}{\gamma_{\rm exc;0}} \times \frac{q}{q_0}.
    \label{eq:genenh}
\end{equation}
Here the quantum yield, $q$, is the ratio of a radiative decay rate, $\tilde{\Gm}_{\rm rad}$, to a total decay rate \ct[Eq. (2)]{Bharadwaj2007},
\begin{equation}
    q  = \frac{\tilde{\Gm}_{\rm rad}}{\tilde{\Gm}_{\rm rad} + \tilde{\Gm}_{\rm nrad} + (1-q_0)/q_0},
    \label{eq:q}
\end{equation}
with $q_0$ being an {\em intrinsic} quantum yield of the emitter,
 $\tilde{\Gm}_{\rm rad, nrad} = (2\tilde{\Gm}^\parallel_{\rm rad, nrad} + \tilde{\Gm}^\perp_{\rm rad, nrad})/3$
is an orientationally averaged decay rate determined at a fixed dipole radial position by averaging over all possible orientations of a dipole emitter, and \textit{tilde} denotes {\em dimensionless} rates normalized to the radiative decay rate of the emitter in the free space.

Fluorescence is a complex process: the rates $\tilde{\Gm}_{\rm rad}$, $\tilde{\Gm}_{\rm nrad}$, and $\gamma_{\rm exc}$ strongly depend on emitter position and, in particular, on its distance from a dissipative metal component~\cite{Moroz2005,Moroz2005a}. Because emitter-metal surface distance, $d$, is an optimization parameter, we write, for the sake of notation, the rates in the formulas above and below without indicating explicitly any positional dependence. In order to assess $F$, mere radiative decay engineering~\cite[Fig. 1]{Lakowicz2002}
does not suffice - both excitation and emission have to be taken into account. As a matter of fact, 
Eq.~\ref{eq:genenh} tells us that an optimal fluorescence enhancement factor requires a delicate balance between the excitation enhancement and quantum yield quenching. 
Such an optimization, revealing an optimal emitter -- metal surface separation, has been a focus of many theoretical and experimental studies over the last two decades~\cite{Ford1984,Geddes2002,Malicka2003,Tovmachenko2006,Anger2006,Bharadwaj2007,Jiao2014,Arruda2017a,Maccaferri2021}.

\begin{figure}[t!]
    \centering
    \includegraphics[width=3.5in]{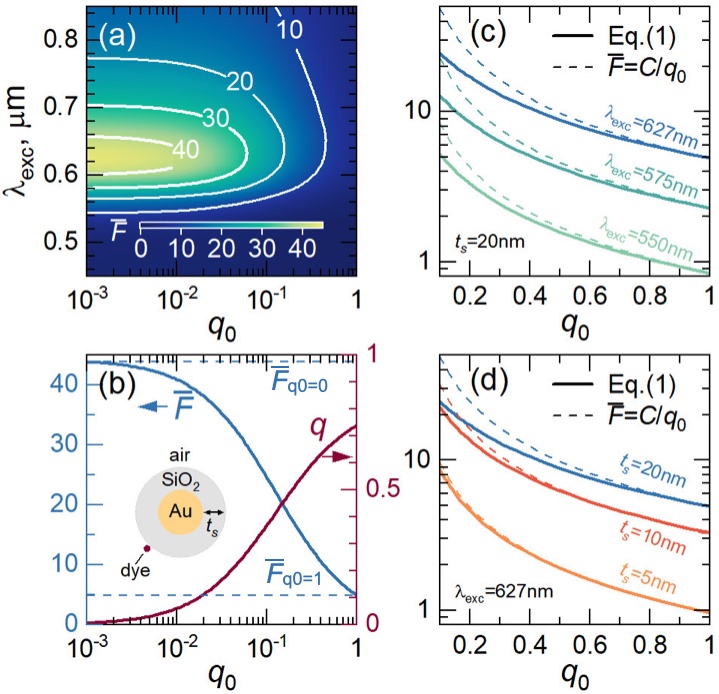}
    \caption{
    (a) Averaged fluorescence enhancement, $\bar{F}$, for Au@SiO$_2$ core-shell nanoparticle as a function of intrinsic quantum yield, $q_0$, and excitation wavelength, $\ld_{\rm exc}$. 
    Core radius is $70$~nm, shell thickness is $20$~nm, dipole emitter is located at $0.75$~nm distance from the shell surface;
    (b) Averaged fluorescence enhancement, $\bar{F}$, and quantum yield, $q$, at $\ld_{\rm exc} = 627$~nm. 
    Dashed horizontal lines show $\bar{F}_{\rm q_0=1}$ and $\bar{F}_{\rm q_0=0}$ according to Eqs.~\ref{eq:Fmin} and \ref{eq:Fmax}, with the excitation rate averaged over the particle surface. Note that high $q$ is not a prerequisite for high $\bar F$.
    (c)-(d) Averaged fluorescence enhancement, $\bar{F}$, calculated with Eq.~\ref{eq:genenh} (solid lines), and with $\bar{F}=C/q_0$ approximation (dashed lines) 
    (c) at $\ld_{\rm exc} = 627,\ 575,\ 550$~nm with $C = 4.9,\ 2.3,\ 0.85$, respectively, and
    (d) at $\ld_{\rm exc} = 627$~nm but for different shell thickness $t_s = 20,\ 10,\ 5$~nm with $C = 4.9,\ 3.2,\ 0.97$, respectively.
    Note the approximate $1/q_0$ scaling of $\bar F$ in (c,d) in the region $(1-q_0)/q_0 \ll \max \{ \tilde{\Gm}_{\rm nrad},\tilde{\Gm}_{\rm rad}\}$. 
    Without losing a generality, an emission wavelength is assumed to be $\ld_{\rm ems} = \ld_{\rm exc}$ in all cases.
    The inset in (b) shows a sketch of a system under consideration.}
    \label{fig:q0}
\end{figure}

It is highly desirable to characterize a plasmonic nanostructure involved in metal-enhanced fluorescence (MEF) with a suitable figure of merit (FOM). The latter would enable one to distinguish potential of different plasmonic platforms for fluorescence enhancement. Purcell factor has often been used to characterize either radiative decay rate or total decay rate changes. It is not suitable as the FOM of a plasmonic nanostructure involved in MEF, because it neglects any influence on $\gamma_{\rm exc}$. 
Alternative to the Purcell factor, the measured or calculated maximum value ${\mathcal F}$ of $F$, where the maximum has been typically determined over predetermined ranges of 
$\ld_{\rm exc}$, $\ld_{\rm ems}$, distances $d$, and relevant structure parameters of a system under consideration, has usually been adopted to characterize considered plasmonic nanostructure. To put it simply, the larger (measured or calculated) ${\mathcal F}$, the more exciting the results~\cite{Kinkhabwala2009,Hoang2015,Traverso2021}.
However, contrary to the conventional practice, $F$, an hence also ${\mathcal F}$, strongly depends on emitter's $q_0$ and, as such, cannot be used to unambiguously characterize emitter environment of a given plasmonic nanostructure.
As immediately seen from Eqs.~\ref{eq:genenh}-\ref{eq:q}, a single plasmonic nanostructure under otherwise identical conditions may yield at any given $\ld_{\rm exc}$, $\ld_{\rm ems}$, and $d$ the values of $F$ strongly \textit{depending on the value of the  emitter's intrinsic quantum yield}, $q_0$. The respective limits for $F$ are:
\bea
   F_{q_0=1} &\equiv & \lim_{q_0 \to 1} F 
   = \frac{\gamma_{\rm exc}}{\gamma_{\rm exc;0}} \times \frac{\tilde{\Gm}_{\rm rad}}{\tilde{\Gm}_{\rm rad} + \tilde{\Gm}_{\rm nrad}},
    \label{eq:Fmin}
    \\   F_{q_0=0} &\equiv & \lim_{q_0 \to 0} F =  \frac{\gamma_{\rm exc}}{\gamma_{\rm exc;0}} \times \tilde{\Gm}_{\rm rad},
    \label{eq:Fmax}
\eea
where, unless $\tilde{\Gm}_{\rm nrad}\equiv 0$, always $F_{q_0=1}< F_{q_0=0}$ (see Fig. \ref{fig:q0}). The very same applies obviously also to ${\mathcal F}$.

When setting the goal of finding a proper FOM for a plasmonic structure involved in MEF, the $q_0$-dependence and emitter orientation have to be discounted when characterizing a nanostructure potential for fluorescence enhancement.
In order to achieve our goal, we are thus led to examine in detail the $q_0$-dependence.
The nonradiative decay rate grows as $\tilde{\Gm}_{\rm nrad}\propto d^{-3}$ when the distance $d$ of an emitter from a metal surface is reduced~\cite{Ford1984}. 
Therefore, a typical situation in a proximity to a metal surface is that at least $\tilde \Gamma_{\rm nrad}\gg 1$, and often also $\tilde \Gamma_{\rm rad}\gg 1$. 
At the same, electric field intensity may be significantly enhanced for $d\to 0$, thus boosting the excitation.
For a typical emitter $0.1 \lesssim q_0  \lesssim 1$, and the third term
$(1-q_0)/q_0$ in the denominator of $q$ in Eq.~\ref{eq:q} varies within the interval $(0,9)$.
As long as one considers metal-fluorophore separations for which $\tilde \Gamma_{\rm nrad}\gtrsim 10^2$, the term
$(1-q_0)/q_0$ in the denominator of $q$ in Eq.~\ref{eq:q} is negligible and the resulting fluorescence enhancement enjoys, in virtue of Eq.~\ref{eq:genenh}, a {\em scaling} behaviour, $F\propto 1/q_0$.
The approximate $1/q_0$ scaling of $F$ holds in the entire region $(1-q_0)/q_0 \ll \max \{ \tilde{\Gm}_{\rm nrad},\tilde{\Gm}_{\rm rad}\}$. 
The scaling region of $q_0$ can be larger or smaller depending on $\max \{ \tilde{\Gm}_{\rm nrad},\tilde{\Gm}_{\rm rad}\}$.
The closer the emission wavelength to the localized surface plasmon resonance (LSPR), the larger $\tilde\Gm_{\rm nrad}$, implying increasing scaling region of $q_0$.
Alternatively, the smaller emitter -- metal surface separation, the larger $\tilde\Gm_{\rm nrad}$, which again increases the scaling region of $q_0$.

We demonstrate the above trends in Fig.~\ref{fig:q0} on a conventional metal-dielectric Au@SiO$_2$ nanoparticle used to enhance a fluorescence of a point-dipole emitter with variable $q_0$ and at different $\ld_{\rm exc}$.
For the sake of generality, the excitation rate in Eq.~\ref{eq:genenh} is \textit{averaged} over the particle surface, $\gamma_{\rm exc} \propto \langle\lvert\vE\rvert^2\rangle$~\cite{Rasskazov19JOSAA,Sun20JPCC}, yielding the \textit{averaged} $\bar F$, which is a reasonable approximation for a typical experiment with spherical particles, unless, of course, the emitter is positioned with a controllable orientation of its electric dipole moment at a particular (e.g. a hot spot) location~\cite{Xin2019}. 
Fig.~\ref{fig:q0} convincingly demonstrates the {\em scaling} behaviour, $\bar F\propto 1/q_0$, {\it i.e.} the smaller $q_0$ the larger $\bar F$, contrary to the main conclusion of \cite{Dragan2012}. Furthermore,
Fig.~\ref{fig:q0} clearly confirms our arguments that the $1/q_0$ scaling holds
for any emitter-metal surface separation in the MEF regime, e.g. irrespective if $\bar F$ attains its maximum or not,
as long as  $(1-q_0)/q_0 \ll \max \{ \tilde{\Gm}_{\rm nrad},\tilde{\Gm}_{\rm rad}\}$.
The closer the emission wavelength to LSPR, or, the smaller emitter -- metal surface separation, the larger $\tilde\Gm_{\rm nrad}$, which increases the scaling region of $q_0$ (Figs.~\ref{fig:q0}(c)-(d)).
The $1/q_0$ scaling holds also for any particular dipole orientation and, consequently, also for averaged dipole orientation. In line with our expectations, a core-shell particle with an emitter with fixed $\ld_{\rm exc}$ may indeed provide dramatically different (by the order of magnitude!) averaged enhancement 
factors: by varying $q_0$ one can change $\bar F$
from $\bar F_{\rm q_0=1}\approx 5$ to $\bar F_{\rm q_0=0}\approx 45$.

To summarize preceding paragraphs, $F$ is {\em ambiguous} when characterizing the environment of an emitter, and hence a plasmonic nanostructure in question. Nevertheless, in spite of all that complexity of fluorescence and its dependence on multitude of variables, such as excitation intensity, emission intensity, directionality and polarization, we were able to verify an approximate scaling, $F\propto 1/q_0$, in the MEF regime. 
Although we provided its explicit verification for a core-shell example, our general theoretical considerations reveal that it is a typical characterizing feature of MEF, irrespective if metal surface is planar, cylindrical, or spherical

To this end, we are going to use this scaling and arrive at the sought emitter's $q_0$ and orientation independent FOM of a plasmonic structure involved in MEF.
Let us pay closer attention to the two limiting values $F_{\rm q_0=0}$ and $F_{\rm q_0=1}$. They are both emitter's $q_0$ and orientation independent, but only the lower limit, $F_{\rm q_0=1}$, depends on both $\tilde{\Gm}_{\rm rad}$ and $\tilde{\Gm}_{\rm nrad}$. 
Therefore, physically, ${\mathcal F}_{\rm q_0=1}$ would provide the {\em worst-case} scenario estimate of ${\mathcal F}$.
For many practical applications, the worst-case estimate alone is a very important quantity enabling one to characterize a device performance. 
At least in the MEF regime, one can recover from ${\mathcal F}_{\rm q_0=1}$ any other realistic ${\mathcal F}$ by a simple $1/q_0$ scaling. Therefore, we advocate here for the use of ${\mathcal F}_{\rm q_0=1}$ as a useful FOM enabling a fair assessment of plasmonic structures for the MEF, see Table~\ref{tab:F}.
For instance, exceptional values of ${\mathcal F}\sim 10^3$ obtained for an organic dye with low $q_0$ ($q_0 = 0.025$~\cite{Kinkhabwala2009} and $q_0=0.1$~\cite{Hoang2015}) reduce to a mere ${\mathcal F}_{\rm q_0=1} \sim 10^2$ after rescaling for dyes with $q_0=1$.
On the other hand, optimally designed plasmonic structures providing $F=910$ for fluorescent dye Atto 532 with $q_0=0.9$~\cite{Traverso2021} are characterized by $F_{\rm q_0=1}=819$. Given ${\mathcal F}_{\rm q_0=1}$ as the FOM, 
local-field corrections \cite{Dolgaleva2012} can, if required, be easily incorporated to arrive at a true fluorescence enhancement.

\begin{table}
\centering
\begin{tabular}{c|c|c|c}
    $F$ & $q_0$ & $F_{q_0=1}$ & Ref. \\
    \hline\hline
    $1340$ & $0.025$ & $\approx33$ & \cite{Kinkhabwala2009} \\
    $2300$ & $0.1$ & $\approx230$ & \cite{Hoang2015} \\
    $910$ & $0.9$ & $\approx819$ & \cite{Traverso2021}
\end{tabular}
\caption{
\label{tab:F}
Comparison between $F_{q_0=1}$ for experimentally measured fluorescence enhancement factors $F$ obtained for dyes with low and large $q_0$.}
\end{table}

Obviously, the scaling cannot continue beyond $q_0\le F_{\rm q_0=1}/F_{\rm q_0=0}$. Then a suitable FOM of a plasmonic nanostructure can be obtained by the pair of two limiting values ${\mathcal F}_{\rm q_0=0}$ and ${\mathcal F}_{\rm q_0=1}$ of Eqs.
~\ref{eq:Fmin} and \ref{eq:Fmax}, where ${\mathcal F}_{\rm q_0=0}$ enables one to set an upper bound on the {\em best-case} scenario. 
Additionally, from Eqs.~\ref{eq:Fmin}--\ref{eq:Fmax} one can immediately infer the importance of tailoring the \textit{excitation} enhancement rather than the \textit{emission} enhancement, keeping $\tilde{\Gm}_{\rm nrad}$ as small as possible for achieving the largest ${\mathcal F}_{\rm q_0=1}$~\cite{Rasskazov21JPCL}. 
Finally, the reported discussion has implications for other anti-Stokes processes. For example, enhancement factor of plasmon-enhanced upconversion~\cite{Wu2014,Qin2021} follows essentially the same behaviour, with the only difference being that the upconversion excitation rate is proportional to the square of the intensity of the electric field, $\gamma_{\rm exc}$ ($\propto \lvert \vE\rvert^4$). Because of quartic field dependence in the latter case, the
local-field corrections may play a more significant role \cite{Dolgaleva2012}.

\section*{Declarations}

{\bf Funding} No funding was received to assist with the preparation of this manuscript.\\
{\bf Conflicts of interest/Competing interests} The authors have no relevant financial or non-financial interests to disclose.\\
{\bf Availability of data} The datasets generated during and/or analysed during the current study are available from the corresponding author on reasonable request.\\
{\bf Code availability} The computer codes used during the current study are available from the corresponding author on reasonable request.\\
{\bf Author Contribution} I.L.R. and A.M. contributed equally to this work.\\
{\bf Ethics approval} Not applicable.\\
{\bf Consent to participate} Not applicable.\\
{\bf Consent for publication} Not applicable.



\end{document}